\documentclass{aa}

\usepackage{graphicx}
\usepackage{txfonts}
\usepackage{multirow}
\usepackage{url}
\usepackage{xcolor}

\begin{document}

   \title{Filament profiles from WISExSCOS galaxies as probes of the impact of environmental effects}

   \author{V. Bonjean\inst{1,2}
   \and
   N. Aghanim\inst{1}
   \and
   M. Douspis\inst{1}
   \and
   N. Malavasi\inst{1}
   \and
   H. Tanimura\inst{1}
    } 

   \institute{Institut d'Astrophysique Spatiale, CNRS, Université Paris-Sud, Bâtiment 121, Orsay, France\\
   \email{victor.bonjean@ias.u-psud.fr}
    \and
    LERMA, Observatoire de Paris, PSL Research University, CNRS, Sorbonne Universités, UPMC Univ. Paris 06, 75014, Paris, France\\
    }

    \date{Received XXX; accepted XXX}

    \abstract{
The role played by the large-scale structures in the galaxy evolution is not quite well understood yet. In this study, we investigate properties of galaxy in the range $0.1<z<0.3$ from a value-added version of the WISExSCOS catalogue around cosmic filaments detected with DisPerSE. We have fitted a profile of galaxy over-density around cosmic filaments and found a typical radius of $r_\mathrm{m} = 7.5\pm0.2$ Mpc. We have measured an excess of passive galaxies near the filament's spine, higher than the excess of transitioning and active galaxies. We have also detected SFR and $\mathrm{M}_\star$ gradients pointing towards the filament's spine. We have investigated this result and found an $\mathrm{M}_\star$ gradient for each type of galaxies: active, transitioning, and passive, and a positive SFR gradient for passive galaxies. We also link the galaxy properties and the gas content in the Cosmic Web. To do so, we have investigated the quiescent fraction $f_\mathrm{Q}$ profile of galaxies around the cosmic filaments. Based on recent studies about the effect of the gas and of the Cosmic Web on galaxy properties, we have modelled $f_\mathrm{Q}$ with a $\beta$ model of gas pressure. The slope obtained here, $\beta=0.54\pm0.18$, is compatible with the scenario of projected isothermal gas in hydrostatic equilibrium ($\beta=2/3$), and with the profiles of gas fitted in SZ.
}


    \maketitle

\section{Introduction}

The matter distribution in the Universe is very non-linear and very complex, structured into a cosmic web composed of voids, walls, filaments, and nodes \citep[e.g.,][]{zeldovich1970, bond1996, libeskind2018}. Understanding the evolution and the physical properties of the matter around the largest scale structures remains one of the main challenge in observational and theoretical cosmology.

Nowadays, numerical simulations such as Millenium \citep{springel2005}, Illustris-TNG \citep{springel2018}, Horizon-AGN \citep{dubois2014}, or Magneticum \citep{hirschmann2014} allow us to trace and characterise all the matter (dark matter, hot gas, cold gas, galaxies) of the large-scale Universe. Among the different structures of the cosmic web, the objects that are the easiest to detect and characterise are the ones with the highest densities, i.e., the nodes, where the galaxy clusters lay. The average properties of these objects are known \citep[e.g.,][]{kravtsov2012, bykov2015, walker2019}, thanks to numerical simulations, but also to actual observations of galaxies in optical and near-infrared, of hot gas via the Sunyaev-Zel'dovich effect \citep[SZ,][]{sunyaev1970, sunyaev1972} and in the X-rays, and of dark matter via gravitational lensing. As a matter of fact, around these structures, dark matter and hot gas universal density profiles have already been derived \citep[e.g.,][]{nagai2007, arnaud2010, planck_gnfw, bartalucci2017, ghirardini2019}, and a global picture of the distribution and of the properties (such as the star formation rate and the stellar mass) of galaxies that fall into their gravitational potential has already been drawn \citep[e.g.,][]{mahajan2009, mahajan2011, baxter2017, chang2018, adhikari2019, pintos2019}.

While galaxy clusters are relatively easy to detect, study, and characterise, other Cosmic Web structures, like cosmic filaments, are not easily defined because of their low densities. No global picture such as the one drawn for the galaxy clusters has been derived yet, and even the definition of cosmic filaments is still arguable since it depends on the way they are detected. Several methods have been proposed to detect the cosmic filaments, for example Bisous \citep{tempel2016}, DisPerSE \citep{sousbie2011}, NEXUS+ \citep{cautun2013}, or very recently T-ReX (Bonnaire et al., in prep.). Each of them has advantages and disadvandages, making comparison between them really difficult (see \cite{libeskind2018} for a review or Bonnaire et al., in prep.). Despite this, recent studies have used those methods to investigate the physical properties of the matter (Dark Matter, gas, or galaxies) in filaments in numerical simulations \citep[e.g.,][Gal\'arraga et al., in prep.]{colberg2005, dolag2006, aragon2010, cautun2014, gheller2015, gheller2016, martizzi2019, gheller2019}. Indeed, studying the gas is only possible in numerical simulations. The hot gas around cosmic filaments is very hard to detect either in SZ or in X-rays because of its low density and low temperature. It is only accessible in a few exceptional objects such as the galaxy cluster pair A399-A401 \citep[e.g.,][]{fujita1996, sakelliou2004, fujita2008, akamatsu2017, bonjean2018}. Alternatively, some studies have used the stacking around the highest density regions in between galaxy cluster pairs \citep{tanimura2019, degraaff2019} or inside super-clusters \citep{tanimura_sc2019} to detect the densest parts of the gas in the filaments around clusters. A first statistical study of the hot gas using the SZ effect around cosmic filaments is performed in \cite{tanimura_fil}.

Galaxies around cosmic filaments, easier to detect, have started to be extensively studied recently in different surveys: in SDSS in 3D \citep[e.g.,][at $z=0.1$, and $z<0.7$]{martinez2016, chen2017, kuutma2017}, in GAMA in 3D \citep[e.g.,][at $z<0.2$ and $0.03<z<0.25$]{alpaslan2015, alpaslan2016, kraljic2018}, in CFHTLS\footnote{\url{https://www.cfht.hawaii.edu/Science/CFHTLS/}} in 2D \citep[e.g.,][at $0.15<z<0.7$]{sarron2019}, in VIPERS in 3D \citep[e.g.,][at $0.5<z<0.85$]{malavasi2017}, and in COSMOS in 2D \citep[e.g.,][at $0.5<z<0.9$]{laigle2018}. These studies show evidence of galaxy population segregation inside filaments, hints of pre-processing and quenching processes of galaxies while entering the large scale structures, and of a positive stellar mass gradient pointing towards the filament's spines. Although these trends are detected in different studies, the mechanisms responsible for these processes are not understood yet, and the role of the environment in evolution of galaxies is still not clear.

In this study, we present a statistical study of galaxy properties from the value-added catalogue based on the WISExSCOS catalogue, around cosmic filaments at low redshift (in the range $0.1<z<0.3$) extracted with DisPerSE in a spectroscopic sample of galaxies from the SDSS. We show the statistical distributions of all galaxies, and of passive, transitioning, and active galaxies around the cosmic filaments, together with their stellar mass and SFR profiles. We also investigate the role of the Cosmic Web and of the hot gas on the galaxy quenching around the cosmic filaments, by linking the quiescent fraction profile to a profile of hot gas. In Sect.~\ref{sect:data}, we present the data we used for the study, i.e., the galaxy density maps based on the WISExSCOS catalogue, the galaxy cluster catalogues used to mask the maps from their emissions, and the catalogue of filaments extracted with DisPerSE from the SDSS galaxies. In Sect.~\ref{chapt:filaments;sect:profiles}, we explain the methodologies used to extract the galaxy over-density profiles around the filaments. In Sect.~\ref{chapt:filaments;sect:results}, we show the different results obtained in matter of galaxy over-density profiles and galaxy properties around filaments, split over three bins of filament lengths. In Sect.~\ref{chapt:filaments;sect:quenching}, we present a tentative to link the profiles of quenched galaxies and the profiles of gas content inside filaments. We summarise the results in Sect.~\ref{chapt:filaments;sect:conclusion}. We assume throughout this paper the \textit{Planck} 2015 cosmology \citep{planck_cosmo2016}, with $\mathrm{H}_0=67.74$ $\mathrm{km/Mpc/s}$, ${\Omega_\mathrm{M}}_0=0.3075$, and ${\Omega_\mathrm{b}}_0=0.0486$.

\section{Data}\label{sect:data}

In this section, we present the data we used to perform the analysis. We first present the WISExSCOS photometric redshift catalogue of galaxies, and then the catalogue of cosmic filaments we based our study on. Finally, we present the combination of galaxy cluster catalogues used to remove the cluster galaxy members.

\subsection{WISExSCOS value-added catalogue and maps}\label{chapt:sfr_mstar;sect:maps}

\subsubsection{WISExSCOS value-added catalogue}\label{chapt:sfr_mstar;sect:vac}

\begin{figure*}[!ht]
\centering
\includegraphics[width=0.5\textwidth]{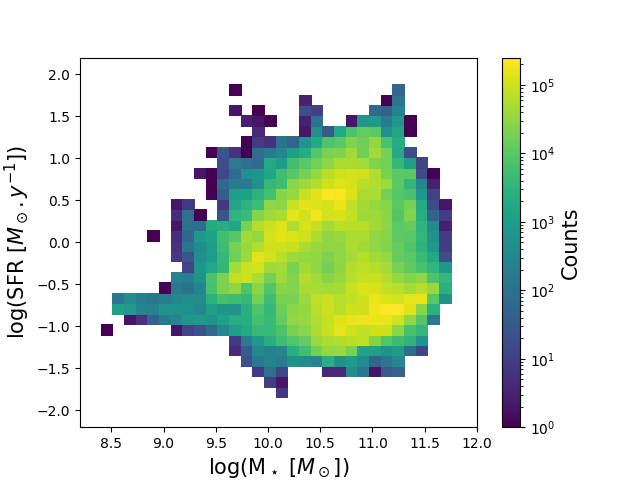}\includegraphics[width=0.485\textwidth]{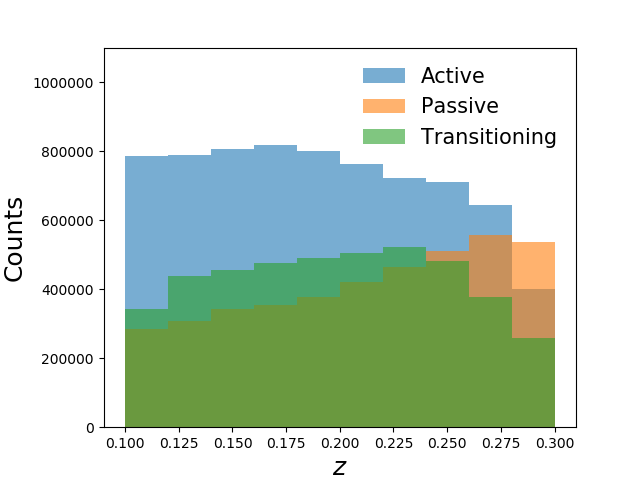}
\caption{\label{chapt:sfr_mstar;histo_gal}Left: range of SFR and of $\mathrm{M}_\star$ of the 15,765,535 sources in the WISExSCOS value-added catalogue in the range $0.1<z<0.3$. Right: distributions of the active, transitioning, and passive galaxies of the WISExSCOS value-added catalogue as a function of redshift.}
\end{figure*}

We based our study on the WISExSCOS photometric redshift catalogue \citep{bilicki2016}. This catalogue contains both photometric redshift estimates and WISE magnitudes. Based on these properties, we have estimate SFR and $\mathrm{M}_\star$ following \cite{bonjean2019}, for 15,765,535 sources in the redshift range $0.1 < z < 0.3$. The range of SFR and of $\mathrm{M}_\star$ of the WISExSCOS value-added catalogue is shown in the left panel of Fig.~\ref{chapt:sfr_mstar;histo_gal}. 

\subsubsection{Distance to the main sequence estimation}\label{chapt:sfr;sect:d2ms}

Similarly to estimating the star-forming activity of galaxies by computing specific SFR (which illustrates the efficiency of a galaxy in forming stars), the distance to the main sequence on an SFR-$\mathrm{M}_\star$ diagram informs us about the star formation activity. This quantity, noted $d2ms$, is estimated following \cite{bonjean2019}, and can be a useful property to segregate populations of galaxies. We have thus splited the 15,765,535 sources of the WISExSCOS catalogue in the range $0.1 < z < 0.3$, into active, transitioning, and passive galaxies using the $d2ms$ criterion. We have defined as active galaxies sources with $d2ms$ < 0.4, as transitioning galaxies sources with 0.4 < $d2ms$ < 1.25, and as passive galaxies sources with $d2ms$ > 1.25. These cuts are defined by the intersections of three Gaussians fitted on the distribution of the $d2ms$ on a sample of spectroscopic SDSS galaxies, used to model the three populations of galaxies. After the splitting, the catalogue contains 7,249,961 active, 4,353,744 transitioning, and 4,161,830 passive galaxies. The distributions of the three populations of galaxies in redshifts are shown in the right panel of Fig.~\ref{chapt:sfr_mstar;histo_gal}.

\subsubsection{Galaxy density maps}\label{chapt:filaments;sect:maps}

\begin{figure*}[!ht]
\centering
\includegraphics[width=\textwidth]{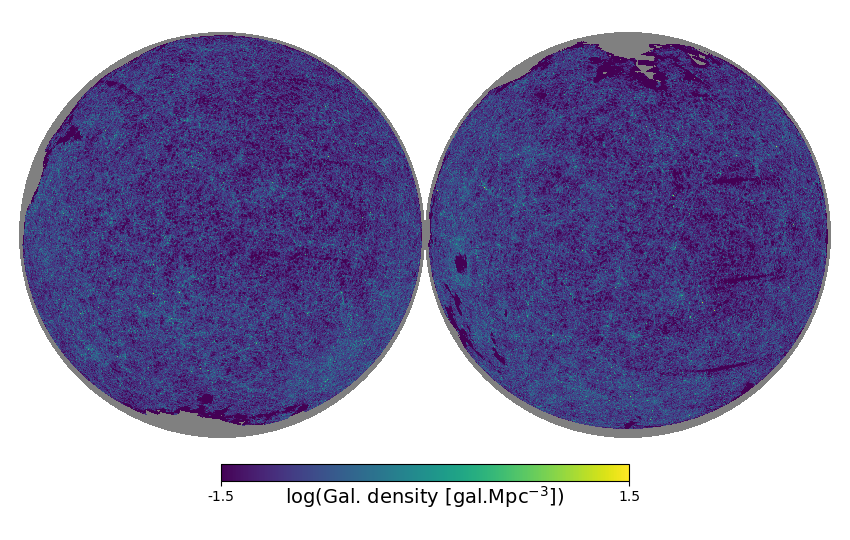}
\caption{\label{chapt:sfr_mstar;example_map}Mollweide projection of the slice at $z=0.15$ of the 3D passive galaxy density map constructed with the pyDTFE code. The map is smoothed at 30' for visualisation. The large-scale distribution of the galaxies is seen, together with artefacts from the WISE scanning strategy, and the masks of our galaxy and of the Magellanic cloud.}
\end{figure*}

Based on the WISExSCOS value-added catalogue and on its three sub-samples of galaxy types defined above, we have constructed 3D galaxy density maps, in the redshift range $0.1<z<0.3$, using the positions of the sources and their redshift information. To do so, we have reconstructed the density field with a python implementation\footnote{\url{https://github.com/vicbonj/pydtfe}} of the Delaunay Tesselation Field Estimator \citep[DTFE,][]{schaap2000}. Based on 3072 3D density fields in patches of $3.7^\circ\times3.7^\circ$, we have generated four set of 20 set of HEALPIX full-sky maps (20 for the 20 bins in redshift): one for all galaxies, and one for each of the three populations of galaxies. We set $n_{\mathrm{side}}=2048$ to the HEALPIX maps so that the pixel resolution is 1.7', and the binning in redshift was arbitrarily set to $\delta_z=0.01$. An example of a slice at $z=0.15$ of the 3D passive galaxy density map (smoothed at 30' for visualisation) is shown in Fig.~\ref{chapt:sfr_mstar;example_map}. The large scale distribution of the galaxies is seen, together with contaminations, i.e., the stripes due to the WISE scanning strategy, the mask of our galaxy and of the Magellanic cloud, and the reddening from dust around our galaxy and the Magellanic cloud. High density concentrations are also seen, which are galaxy clusters. In addition to the four galaxy density maps, we have constructed in the same way 3D maps of SFR and $\mathrm{M}_*$ for all galaxies, and for the active, transitioning, and passive populations. The maps are constructed by linearly interpolating at their 3D positions the estimated SFR and $\mathrm{M}_*$ quantities.

\subsection{Catalogue of filaments}\label{sect_data2}

We have used the SDSS-Data Release 12 (DR12) LOWZCMASS spectroscopic sample of galaxies to construct a catalogue of cosmic filaments using DisPerSE \citep{sousbie2011}. DisPerSE is an algorithm that detects filaments in catalogues of discrete points. The method is based on the density field, reconstructed with the DTFE method. DisPerSE identifies in the density field critical points, i.e., points of maximum density, saddle points, bifurcation points, and points of minimum local density. It then outputs a catalogue of filaments, as connecting the saddle points to the maximum density points. Each filament is constituted by several segments, with given positions and redshifts. The persistence of the filaments, related to their significance, was set to 3 $\sigma$. The details of the construction of the catalogue are given in \cite{malavasi2019}. We have defined the mean positions of the filaments ($\mathrm{R.A.}_\mathrm{mean}, \mathrm{Dec.}_\mathrm{mean}$) being the mean of the positions ($\mathrm{R.A.}_i, \mathrm{Dec.}_i$) of the $i$ segments. In the following, we have defined the minimum, mean, and maximum redshifts of the filaments, $z_\mathrm{min}$, $z_\mathrm{mean}$, and $z_\mathrm{max}$, as the minimum, the mean, and the maximum redshifts of the segments composing the filaments.

\begin{figure*}[!ht]
\centering
\includegraphics[width=0.5\textwidth]{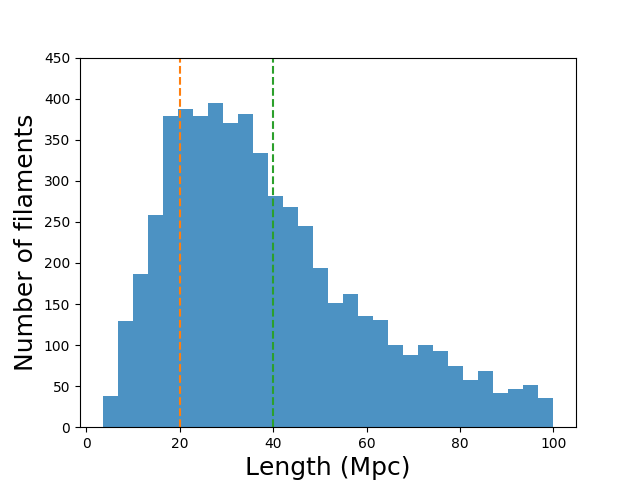}\includegraphics[width=0.5\textwidth]{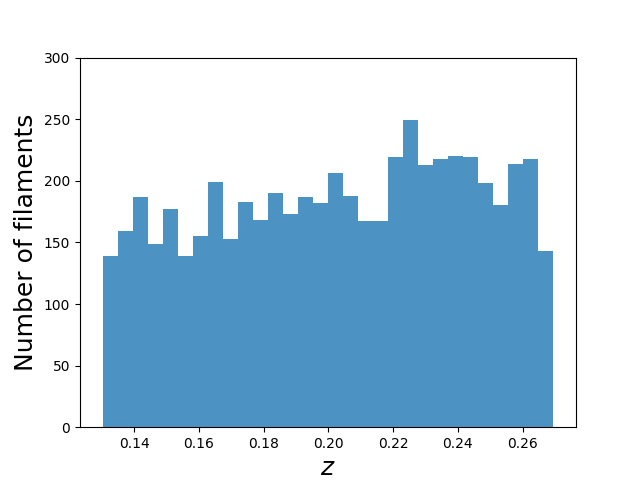}
\caption{\label{chapt:filaments;histo_fil}Selection of the SDSS-DR12 LOWZCMASS DisPerSE filaments used here. Left: 3D length distribution. The orange dotted line shows the limit between the short and the regular filaments, and the green dotted line shows the limit between the regular and the long filaments. Right: redshift distribution.}
\end{figure*}

Since we study the properties of the WISExSCOS value-added catalogue, we have thus excluded the filaments for which parts went outside the redshift range of the WISExSCOS catalogue, that is $0.1<z<0.3$, to ensure that the filaments of our selection are entirely studied. We have also cut the longest filaments at $l>100$ Mpc (where $l$ is the 3D length of the filament in Mpc), that may be non reliable filaments. We have also excluded the filaments for which profiles were not complete so that each bin in profile has the same statistics. A final selection of $n=5559$ cosmic filaments is used in our analysis. Their length and redshift distributions are shown in Fig.~\ref{chapt:filaments;histo_fil}.

We have split the 5559 selected DisPerSE filaments over three bins defined by their 3D length $l$ (in Mpc), to later study the effect of the environment on galaxy properties around the filaments. In the following, we define ``short'', ``regular'', and ``long'' filaments, as follows:

\begin{itemize}
\item $3<l<20: n_\mathrm{obj} = 1042$,
\item $20<l<40: n_\mathrm{obj} = 2291$,
\item $40<l<100: n_\mathrm{obj} = 2226$.
\end{itemize}

For the case of short filaments, we have set the upper limit to 20 Mpc based on the information given by the 2-point correlation functions of groups in the 2DF survey \citep{yang2005}, that infer that filaments below this typical size may be the tiniest and densest filaments, i.e., bridges of matter connecting two clusters, such as the bridge between A399 and A401 \citep{bonjean2018}. We have set the limit between the regular and the long filaments arbitrarily to 40 Mpc to keep the same statistics in the two categories (about 2250 filaments in each).

\subsection{Galaxy cluster catalogues}\label{sect_data3}

We have masked the over-density maps at the position of known clusters, preventing us from including galaxy cluster members while we compute the galaxy over-density profiles around the filaments. We use for the masking a combination of catalogues of galaxy cluster detected in X-ray, optical, and via the SZ effect. We masked 1,478 clusters from the 2015 \textit{Planck} PSZ2 clusters catalogue \citep{planck_psz22016}, 1,601 clusters from the MCXC ROSAT X-ray clusters catalogue \citep{piffaretti2011}, 15,846 clusters from the galaxy clusters detected in optical in the Sloan Digital Sky Survey \citep[SDSS,][]{york2000} from RedMapper \citep{rykoff2014}, 2,395 from WHL 2015 \citep{wen2015}, 75,045 from WHL 2012 \citep{wen2012}, and 33,428 from AMF9 \citep{banerjee2018}. We masked only galaxy clusters with $z<0.4$ to ensure, in a conservative way, that most of the galaxy clusters in the redshift range $0.1<z<0.3$ of the WISExSCOS catalogue are taken into account. In order to be very conservative, we also have masked the catalogue of bifurcation points and of maximum density points with $z<0.4$ provided by DisPerSE, as these points can be at positions of groups or clusters of galaxies that may not be detected in the galaxy cluster catalogues. We investigate the use of different masks in Sect.~\ref{chapt:filaments;sect:mask}, where different radii for masking clusters have been used.

\section{Measuring the profiles}\label{chapt:filaments;sect:profiles}

\subsection{Methodology}

In order to measure the radial profiles of galaxy quantities around filaments, we have used the RadFil code developed by \cite{zucker2018}. RadFil takes as input two maps: one of the quantity of interest, here the maps of galaxy densities constructed in Sect.~\ref{chapt:sfr_mstar;sect:maps}, and one tracing the spine of the filament around which it will perform the measurement, here the maps tracing the spines of the selected DisPerSE filaments. Different steps were needed in order to obtain the radial profiles; they are explained hereafter.

\begin{itemize}
\item Normalisation of the maps

The galaxies in the WISExSCOS value-added catalogue are not uniformly distributed in redshift, and the three population of galaxies do not follow the same distributions, as it is shown in Fig.~\ref{chapt:sfr_mstar;histo_gal}. Thus, the values of the mean galaxy densities in each redshift slice of the galaxy density maps constructed in Chap.~\ref{chapt:sfr_mstar;sect:maps} also follow the same redshift distribution. Measuring the absolute value of the galaxy densities may thus introduce bias. To avoid this, and to measure only the excess of galaxies relative to the mean galaxy density in the field, we have normalised the 3D density maps in order to consider over-densities $\delta$. We have divided each slice of redshift of each 3D map by their mean galaxy density values:

\begin{equation}
1+\delta_\mathrm{gal}(z) = \frac{\rho_\mathrm{gal}(z)}{<\rho_\mathrm{gal}(z)>}.
\end{equation}

In that way, the 3D maps are transformed from biased densities $\rho_\mathrm{gal}$ to unbiased over-densities $1+\delta_\mathrm{gal}$.

\item Projection on patches

For each of the 5559 filaments, we have projected the 3D maps of the obtained galaxy over-density $1+\delta_\mathrm{gal}$, on 3D patches centred on the position of the corresponding filament using a tangential projection. The 3D patches have a pixel resolution of $\theta_\mathrm{pix}=1.7$ arcmin, a bin in redshift of 0.01 (same as the full-sky maps), and a number of pixels which depends on the length of the filament (computed in arcmin with the mean redshift $z_\mathrm{mean}$). Doing so, all filaments are entirely seen in their corresponding individual patch and the fields of view of the largest patch is $19^\circ\times19^\circ$. As 95\% of the patches have fields of view below $15^\circ\times15^\circ$, we have neglected the projection effects and assumed the flat sky approximation.

\item Stack along redshift

Due to the high value of the statistical error on the photometric redshifts of the sources in the WISExSCOS catalogue, $\sigma_z=0.033$, 3D density profiles around filaments would be biased. Therefore, we have stacked the 3D patches (obtained above) along the redshift axis to remove the uncertainty on the positions of the galaxies in the redshift space. The resulting stacked maps are thus 2D arrays. Before stacking along redshift, in order to minimise the noise due to background and foreground galaxies, we have removed for each filament the regions of the 3D patch that lie outside the redshift range $z_\mathrm{min}-\sigma_z < z < z_\mathrm{max}+\sigma_z$, where $z_\mathrm{min}$ and $z_\mathrm{max}$ are the minimum and maximum redshifts of the filament. Mathematically, this step translates into:

\begin{equation}
<1+\delta_\mathrm{gal}> = \frac{1}{b_z}\sum_{z=z_\mathrm{min}-\sigma_z}^{z_\mathrm{max}+\sigma_z} \left(1+\delta_\mathrm{gal}(z)\right),
\end{equation}

where $b_z$ is the number of redshift bins in the range [$z_\mathrm{min}-\sigma_z$, $z_\mathrm{max}+\sigma_z$].

\item Application of RadFil

We have fed RadFil with the 5559 2D stacked maps obtained above, together with the 5559 associated 2D filament's spine projections, also in the format of 2D arrays. RadFil then measures the radial profiles $<1+\delta_\mathrm{gal}>(r)$ around each of the 5559 filaments.

\item Stack the profiles

Finally, we have stacked the 5559 profiles to get one unique profile, exhibiting statistical trends thanks to the significant reduction of the noise.

\item Estimation of the error bars

To estimate the errorbars on the stacked profiles, we have used the bootstrap method that resamples to ensure the significance of the detection. For the $n$ measured profiles (where $n$ is the number of filaments), we have randomly selected $n$ over $n$ profiles with replacement, and have computed the mean profile. We have repeated this measurement 1000 times and have computed the mean and the standard deviation of the 1000 mean profiles. The mean and standard deviations are taken as final measurements and errors in this study.

\end{itemize}

\subsection{Masking the galaxy cluster members}\label{chapt:filaments;sect:mask}

In order to measure galaxy over-density profiles along filaments uncontaminated by the galaxy cluster members, we have removed the regions around known galaxy clusters, by masking the maps at the position of the clusters presented in Sect.~\ref{sect_data3}.

To mask the clusters in an optimal way, we have defined six masks: the first one where the galaxy clusters are not masked, and the five others where the clusters were masked in regions from $1\times\mathrm{R}_{500}$ to $5\times\mathrm{R}_{500}$. The results are shown in Fig.~\ref{chapt:filaments;compare_r500}. The galaxy over-density profiles decrease with increasing radius of the mask, up to $r=3\times\mathrm{R}_{500}$. Beyond this radius, the profiles are unchanged but the error-bars increase. We have thus chosen to mask the clusters at $r=3\times\mathrm{R}_{500}$. For clusters without estimated $\mathrm{R}_{500}$ (only a handful from the \textit{Planck} catalogue in the SDSS area), we have masked a region with radii increasing up to $r=10$ arcmin. We show in Fig.~\ref{chapt:filaments;compare_arcmin} that masking at $r=5$ arcmin is enough, as no difference in the profiles is noticed. We have also masked regions around the critical points provided by DisPerSE, namely the maxima density points, and the bifurcation points, with $z<0.4$. For these regions, we have masked regions with areas defined by radii varying up to $r=20$ arcmin (see Fig.~\ref{chapt:filaments;compare_crit}). For the critical points, the profiles are decreasing with the increasing radii up to 20 arcmin. Part of the reason for the decrease of the profiles is due to the loss of the shortest filaments in the selection (i.e., a change in the length distribution seen in the left panels). We have chosen to mask the critical points from DisPerSE at $r=10$ arcmin, as it is a good compromise to keep the shortest filaments of about $l<7$ Mpc in the selection of filaments while removing contamination from galaxies in small groups.

\begin{figure}[!t]
\centering
\includegraphics[width=0.5\textwidth]{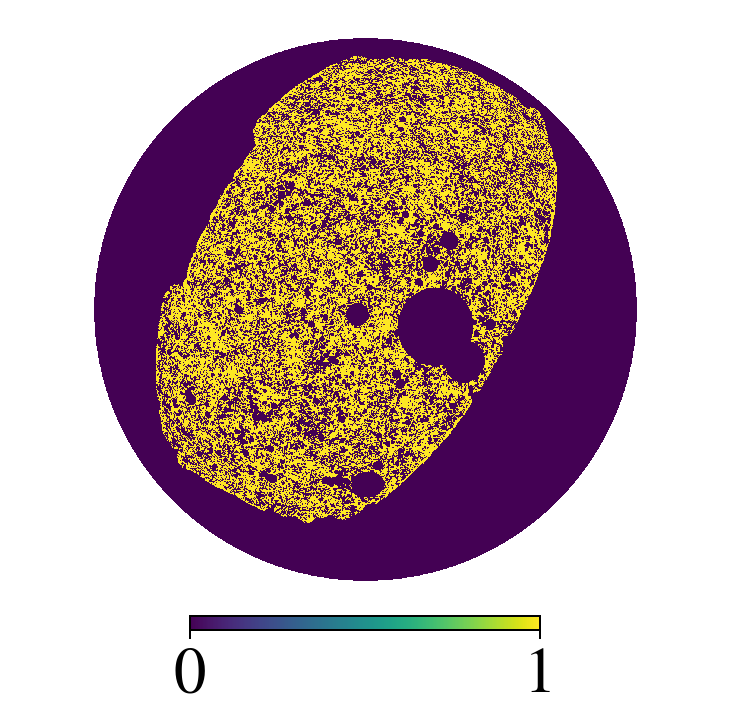}
\caption{\label{chapt:filaments;coverage}Orthographic projection of the union mask of the northern hemisphere used in this analysis.}
\end{figure}

Finally, we have also masked the area outside the footprint of the SDSS-DR12 area. The union mask is shown in orthographic projection in Fig.~\ref{chapt:filaments;coverage} (only the northern hemisphere is shown, as none of the selected filaments are in the southern hemisphere).

\subsection{Robustness of the measurement}\label{chapt:filaments;sect:measure}

To ensure the robustness of the measurements and compute their significance, we have performed null-tests. By applying the exact same methodology as the one detailed above, we have computed profiles in patches with the same properties as the ones of the 5559 DisPerSE filaments (number of pixels $\theta_\mathrm{pix}$ and redshift range $z_\mathrm{min}-\sigma_z < z < z_\mathrm{max}+\sigma_z$), but centred on random positions on the sky, and with random rotations. For computational time reasons, we have computed only 100 random position stacked profiles. The null-test was found compatible with zero at 1 $\sigma$, with an offset of the order of 1\%. However, this offset is negligible and does not change the interpretation of the observed galaxy over-density profiles around filaments, which were all detected with high significances, ranging between $\sim5$ $\sigma$ and 32 $\sigma$. After testing for the effects of the DTFE, the interpolation on the HEALPIX pixels, and the redshift bins, we found that the offset is due to a combination of the mask and the spatial distribution of galaxies. Indeed, we verified that when we reshuffled the pixels in the footprint, the null-test converged to the expected mean value of the map.

Although the profiles are very significantly detected, there are some effects that may produce biases and trends that may not be explicitly related to galaxies around filaments. We discuss the possible sources of contamination hereafter.

One of the main issues is the error on the photometric redshifts of the sources in the WISExSCOS catalogue, $\sigma_z=0.033$. This error may induce a bias during the construction of the 3D density maps presented in Chap.~\ref{chapt:sfr_mstar;sect:maps} and make them not reliable in the redshift direction. To mitigate this effect, we have computed the profiles in 2D projected maps stacked in the range [$z_\mathrm{min}-\sigma_z$, $z_\mathrm{max}+\sigma_z$], where $z_\mathrm{min}$ and $z_\mathrm{max}$ are the minimum and maximum redshifts of the filaments. 

Another possible source of bias may be the 3D to 2D projection of the over-densities. RadFil computes profiles on 2D while the filaments are defined as 3D structures. This may induce a bias in the shape of the profiles, especially when the axis of the filaments are not orthogonal to the line of sight. A comparison with numerical simulation should be done.

Another issue is the reliability of the filaments detected with DisPerSE. Indeed, although the LOWZCMASS sample was chosen to minimise these effects, the SDSS galaxies used for the detection may still suffer from holes in the spatial distribution due to foreground bright stars, biased distribution in redshift due to magnitude limit selection, etc. The filament's positions and redshifts may also be affected by the noise of the survey, and by the finger-of-god effect of the galaxies in clusters \citep[e.g.,][]{jackson1972}, producing shifts of the galaxies in redshift space. The strategy adopted in this study to estimate over-density profiles of galaxies on a catalogue of galaxies (WISExSCOS) independent of the one used to detect the filaments (SDSS) should reduce these biases, since the noises and the biases of the two galaxy catalogues are independent.

\section{Properties around cosmic filaments}\label{chapt:filaments;sect:results}

\subsection{Average properties}\label{chapt:filaments;sect:universal}

\begin{figure}[!ht]
\centering
\includegraphics[width=0.5\textwidth]{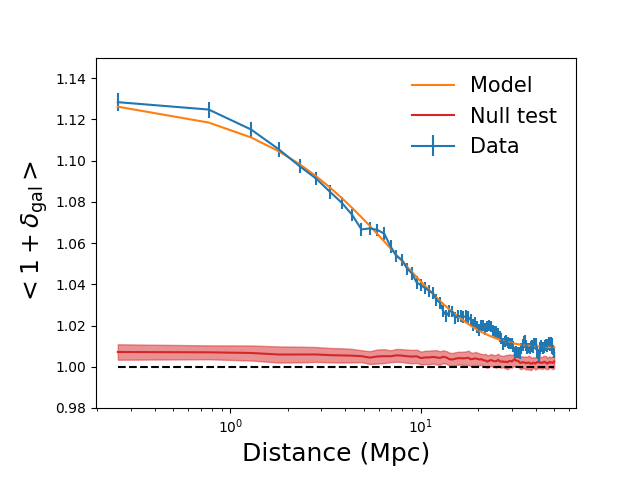}
\caption{\label{chapt:filaments;universal}In blue, the stacked radial profile of over-density $<1+\delta_\mathrm{gal}>$ of the galaxies from WISExSCOS catalogue around the 5559 filaments. In orange, the best fitted exponential model presented in Eq.~\ref{chapt:filaments;eq:exp}. In red, the 100 measurements on random positions from the null tests.}
\end{figure}

The stacked profile $<1+\delta_\mathrm{gal}>$ of the 5559 filaments selected from the DisPerSE catalogue is shown in blue in Fig.~\ref{chapt:filaments;universal}. The red lines and errors show the results of 100 null-tests measurements on random positions and rotations. The binning in distance to the filament of the profiles is from $r=0.25$ to $r=50$ Mpc. Note that due to the pixel resolution of the galaxy density maps $\theta_\mathrm{pix}$, no information with a distance $r<0.25$ Mpc is available. Profiles presented in this study thus show the behaviour of galaxies at large distances from the filament's spines. We have modelled the galaxy over-density profile by an exponential law (show in orange in Fig.~\ref{chapt:filaments;universal}), that takes the form:

\begin{equation}\label{chapt:filaments;eq:exp}
\delta_\mathrm{gal}(r) = {\delta_\mathrm{gal}}_0 \times e^{-\frac{r}{r_\mathrm{m}}}+c_\mathrm{gal},
\end{equation}

where ${\delta_\mathrm{gal}}_0$ is the mean projected over-density in the spines, $r_\mathrm{m}$ the typical radius, and $c_\mathrm{gal}$ is the offset discussed in Sect.~\ref{chapt:filaments;sect:measure} that is related to the interplay of the mask shape and the underlying galaxy distribution which depends on the size of the filaments (shown later in Fig.~\ref{chapt:filaments;universal}).

The fit of the model to the data with a least square minimisation gives ${\delta_\mathrm{gal}}_0 = 0.121\pm0.002$, $r_\mathrm{m} = 7.5\pm0.2$ Mpc, and a very small offset $c_\mathrm{gal}=0.0093\pm0.0004$.

Although some studies have found that dark matter halo density profiles around filaments follow a power law with index between -2 and -3 beyond 3 Mpc \citep{colberg2005, dolag2006, aragon2010}, density profiles of galaxies around filaments have not been studied much. \cite{gheller2019} have fitted a power law with index -1.84 to galaxy sized halo density profiles around filaments, but as shown in \cite{cautun2014}, filaments are very different in terms of width (with width from 2 Mpc to 10 Mpc for the largest ones), making the average density profiles of galaxies very dependent on the selection of filaments. In addition, the profiles presented here may also contain the correlations between the filaments, that is the probability to cross another filament up to a certain radius, and the probability to cross the same filament another time at another radius if the filament is curved. These effects would results in an enlargement of the profile. Hence, the value of the fitted typical radius $r_\mathrm{m} = 7.5\pm0.2$ Mpc might be slightly over-estimated.

\begin{figure*}[!ht]
\centering
\includegraphics[width=0.5\textwidth]{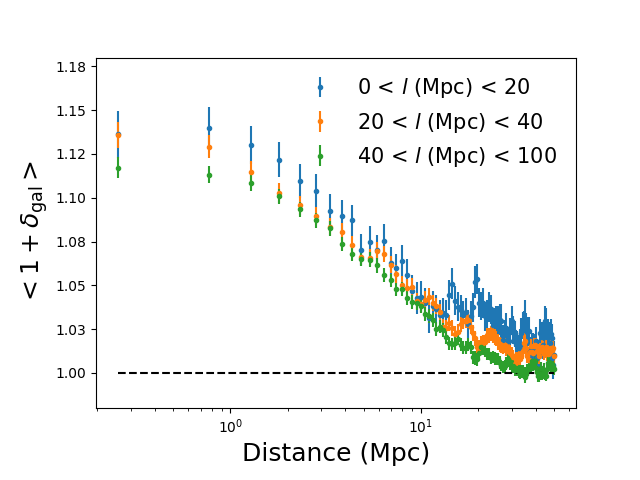}\includegraphics[width=0.5\textwidth]{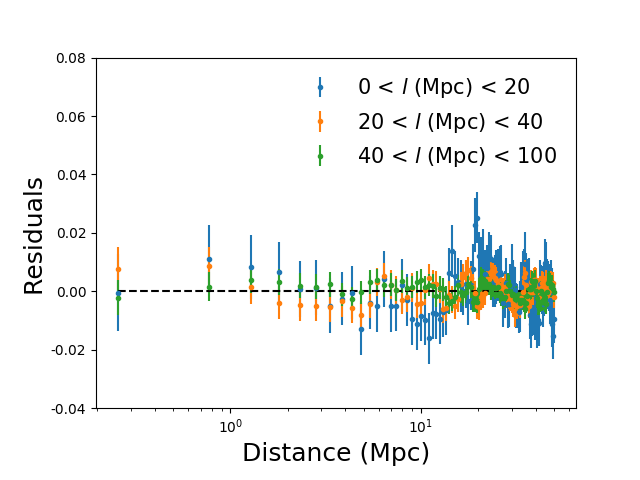}
\caption{\label{chapt:filaments;universal2}Left: radial profiles of galaxy over-densities $<1+\delta_\mathrm{gal}>$ around cosmic filaments, for short, regular, and long filaments. Right: residuals after substracting the exponential model fitted to the average over-density profile, shown in orange in Fig.~\ref{chapt:filaments;universal}, and the tiny offsets.}
\end{figure*}

To study the effect of the global environment, we have averaged the galaxy over-density profiles in the three categories of filaments: the short, the regular, and the long filaments. The resulting profiles are shown in the left panel of Fig.~\ref{chapt:filaments;universal2}. A small offset is seen for the regular filaments ($c_\mathrm{gal}=0.013$) and for the short filaments ($c_\mathrm{gal}=0.018$). However, all three average profiles share the same shape. This trend is confirmed by the computation of the residuals of the three profiles with the exponential model (Eq.~\ref{chapt:filaments;eq:exp}), with the parameters from the average profile (${\delta_\mathrm{gal}}_0 = 0.121\pm0.002$ and $r_\mathrm{m} = 7.5\pm0.2$ Mpc). The residuals are shown in the right panel of Fig.~\ref{chapt:filaments;universal2}. They do not show significant bias outside the error bars, suggesting that the galaxies around the small, regular, and long filaments, are sharing the same shapes, or that the errors are too large to see a significant difference.

\begin{figure*}[!ht]
\centering
\includegraphics[width=0.5\textwidth]{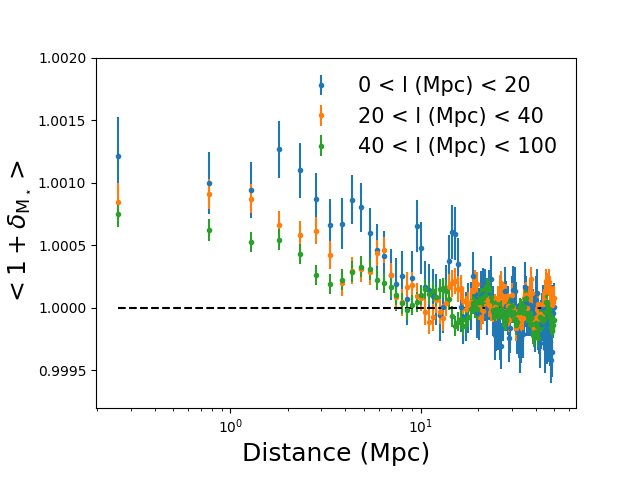}\includegraphics[width=0.5\textwidth]{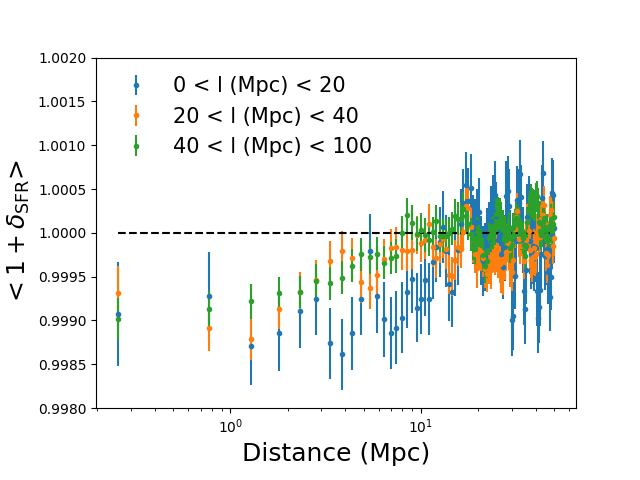}
\caption{\label{chapt:filaments;prop}Radial stacked profiles of excess of $\mathrm{M}_*$ and SFR: $<1+\delta_{\mathrm{SFR}}$> and $<1+\delta_{\mathrm{M}_\star}>$, for the short, the regular, and the long filaments.}
\end{figure*}

We have also computed the stacked profiles of excess of SFR and excess of $\mathrm{M}_*$, defined as $<1+\delta_{\mathrm{SFR}}$> and $<1+\delta_{\mathrm{M}_\star}>$ in the same way as for $<1+\delta_\mathrm{gal}>$. The resulting stacked profiles for the short, the regular, and the long filaments are shown in Fig.~\ref{chapt:filaments;prop}. Gradients of SFR and $\mathrm{M}_\star$ pointing towards the spines of the filaments are detected with a significance of 6.5 $\sigma$ and 9.5 $\sigma$, respectively. Galaxies are significantly $\sim10$\% more massive, and $\sim10$\% less star-forming. This trend is expected since the excess of passive galaxies near the filaments spines was already shown in several studies \citep[e.g.,][]{malavasi2017, laigle2018, kraljic2018, sarron2019}. This implies a different ratio of passive over active galaxies with increasing distance to the filaments' spine, producing these gradients. The $\mathrm{M}_\star$ gradient is 60\% higher in the case of short filaments. This trend is also expected, as short filaments may represent bridges of matter, i.e., more mature structures lying in the densest environments of the Cosmic Web \citep[e.g.,][]{aragon2010} or event resulting from cluster interactions. These environments are already showing evidence of hot diffuse gas potentially inducing quenching of galaxies. For example in the bridge of matter between A399 and A401, most of the galaxies between the two clusters were found to be passive \citep{bonjean2018}. Therefore, it is hard to identify the origin of the observed gradients at this stage. They may reflect a change in the galaxy populations, or, a real stellar mass gradient of all types of galaxies around filaments.

\subsection{Spliting over galaxy types}\label{chapt:filaments;sect:gal_types}

\begin{figure*}[!ht]
\centering
\includegraphics[width=0.33\textwidth]{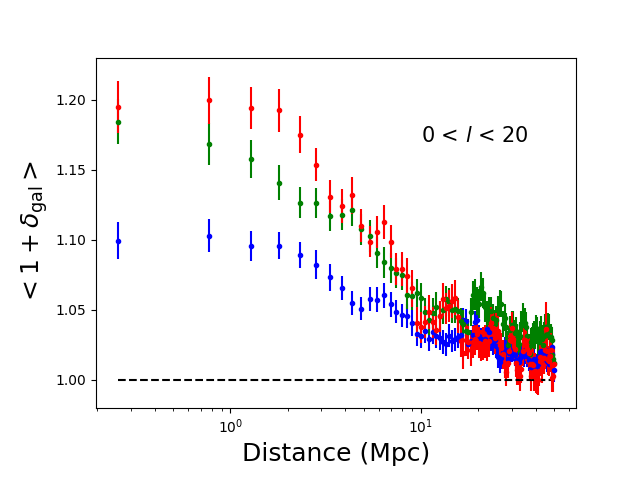}\includegraphics[width=0.33\textwidth]{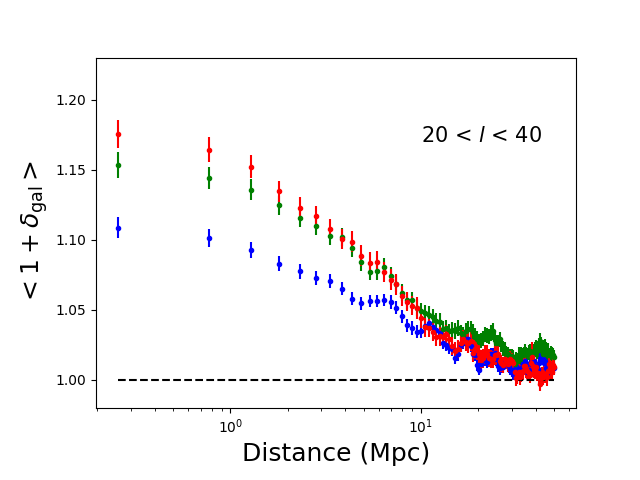}\includegraphics[width=0.33\textwidth]{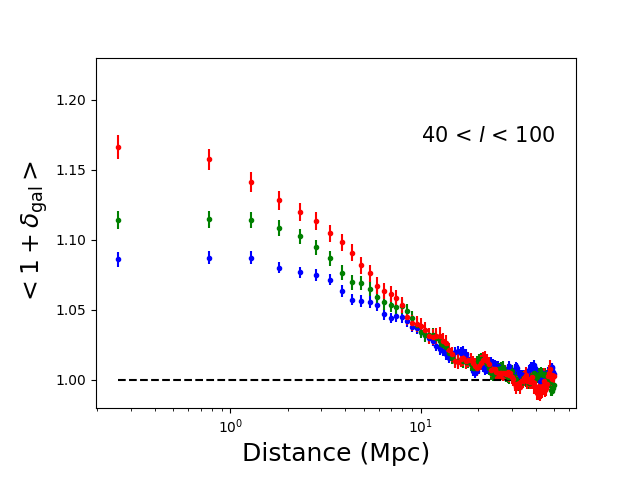}
\caption{\label{chapt:filaments;gal_types}Stacked radial over-density profiles of active (in blue), transitioning (in green), and passive (in red) galaxies, as defined with the distance to the main sequence (detailed in Chap.~\ref{chapt:sfr_mstar;sect:vac}), around the short, the regular, and the long cosmic filaments from left to right.}
\end{figure*}

To explore further the impact of the environment on galaxy properties, we have computed the stacked radial profiles on galaxies, by splitting them in three populations, i.e., active, transitioning, and passive. We have used the 3D maps of galaxy densities constructed for the three types. The resulting stacked profiles $<1+\delta_\mathrm{gal}>$ around the short, regular, and long filaments are shown in Fig.~\ref{chapt:filaments;gal_types}.

The profiles of galaxy over-density are decreasing with the galaxy types: the excess of passive galaxies around the spines of the filaments is higher than the excess of transitioning galaxies, and the excess of transitioning galaxies is higher than the active galaxies. These results show that whatever the length of the filaments, more passive galaxies are distributed towards the spines of the filament. This indicates a quenching process inside the cosmic filaments, as already shown in, e.g., \cite{martinez2016, malavasi2017, kraljic2018, laigle2018, sarron2019}. Another interesting trend to notice is the excess of passive galaxies that is higher around the short filaments than around regular and long ones. This confirms the trend that short filaments may be bridges of matter. However, no difference is seen between the regular and long filaments. These results show that from a length of 20 Mpc, cosmic filaments share on average the same properties and the same galaxy population distributions whatever their length.

\begin{figure*}[!ht]
\centering
\includegraphics[width=0.5\textwidth]{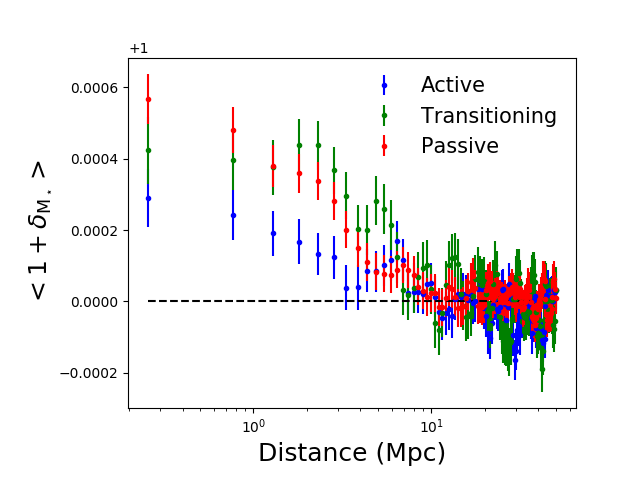}\includegraphics[width=0.5\textwidth]{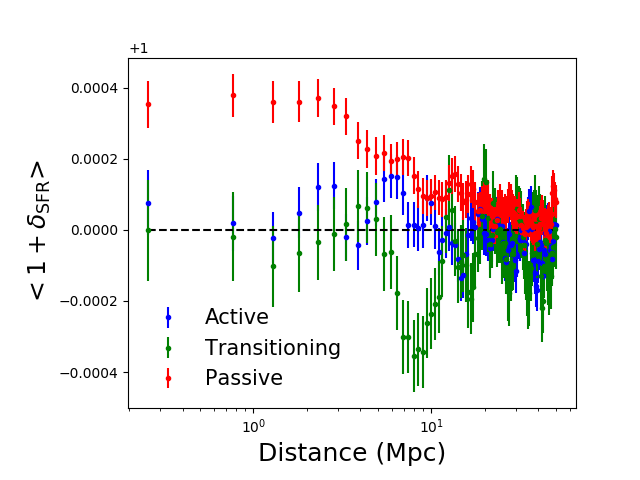}
\caption{\label{chapt:filaments;mass_gradients}Left: $<1+\delta_{\mathrm{M}_\star}>$ stacked radial profiles for each galaxy types: active galaxies in blue, transitioning galaxies in green, and passive galaxies in red. Right: $<1+\delta_\mathrm{SFR}$> stacked radial profiles for the same galaxy types.}
\end{figure*}

We have also computed the excess of $\mathrm{M}_\star$ and SFR for the three galaxy types. The resulting stacked profiles are shown in Fig.~\ref{chapt:filaments;mass_gradients}. In the left panel, $\mathrm{M}_\star$ gradient is seen for each types of galaxies: the active, the transitioning, and the passive ones, detected at 2.2 $\sigma$, 4.3 $\sigma$, and 5.3 $\sigma$, respectively. These results show that the galaxies are $\sim5$\% more massive in the spine of the filaments, whatever the types. An $\mathrm{M}_\star$ gradient towards the spines of the filaments in active galaxies was also detected in \cite{malavasi2017} and in \cite{kraljic2018}. This may be due to mergers of galaxies inside filaments \citep[e.g.,][]{codis2015, malavasi2017}. In the right panel of Fig.~\ref{chapt:filaments;mass_gradients}, a positive SFR gradient is detected at 6.3 $\sigma$, only for passive galaxies. This suggests that passive galaxies are more star-forming inside the filaments. This trend could be due to the rejuvenation of the star formation in passive galaxies, due to gas accreted by galaxy mergers.

\section{A profile derived by the quenching}\label{chapt:filaments;sect:quenching}

\subsection{Quiescent fraction}

Motivated by the results of Sect.~\ref{chapt:filaments;sect:gal_types}, that indicate that galaxy populations are differently distributed towards the spines of the filaments, we have computed the fraction of quenched galaxies, that is similar to the quiescent fraction $f_\mathrm{Q}$ \citep[e.g.,][]{fontana2009, hahn2015, martis2016, lian2016} but with galaxy over-densities rather than galaxy densities. Here the quiescent fraction is defined as the ratio between the excess of passive galaxies $<1+{\delta_\mathrm{gal}}_\mathrm{P}>$ and the sum of the excess of passive $<1+{\delta_\mathrm{gal}}_\mathrm{P}>$ and active $<1+{\delta_\mathrm{gal}}_\mathrm{A}>$ galaxies:

\begin{equation}
f_\mathrm{Q}=\frac{<1+{\delta_\mathrm{gal}}_\mathrm{P}>}{<1+{\delta_\mathrm{gal}}_\mathrm{A}>+<1+{\delta_\mathrm{gal}}_\mathrm{P}>}.
\end{equation}

\begin{figure}[!ht]
\centering
\includegraphics[width=0.5\textwidth]{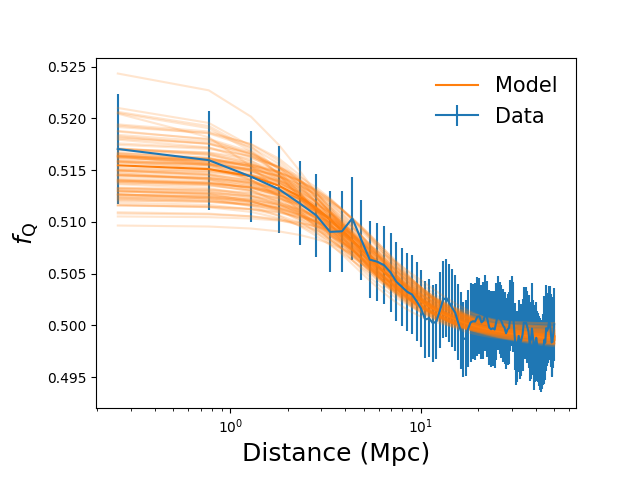}
\caption{\label{chapt:filaments;red_over_blue}In blue, the quiescent fraction profile $f_\mathrm{Q}$, averaged around the 5559 selected filaments. In orange, the $\beta$-model (Eq.~\ref{chapt:filaments;eq:beta}) from different realisations of the MCMC.}
\end{figure}

Since the transitioning galaxies are in the process of quenching and can be defined both as star forming, or quenched, we did not take them into account. The quiescent fraction $f_\mathrm{Q}$ around cosmic filaments is shown in Fig.~\ref{chapt:filaments;red_over_blue}.

\subsection{A link to the gas content?}\label{chapt:filaments;sect:gas}

It is still not clear to date what are the dominant drivers of the quenching process in the galaxies. It has been already shown, and confirmed in the present study, that the Cosmic Web filaments have an impact on galaxy properties \citep[e.g.,][]{malavasi2017, laigle2018, kraljic2018, sarron2019}. Some studies have assumed the quenching being only due to the environment and to the Cosmic Web, via the hot gas in the halos ($\mathrm{T} \geq 10^{5.4}$ K) preventing the galaxies from forming stars \citep{gabor2015} or by Cosmic Web Detachment \citep{aragon2016}. \cite{peng2015} and \cite{trussler2018} have studied the effect of the gas of the Cosmic Web on the quenching of the galaxies, and have concluded that starvation and strangulation were the main processes of quenching. In each of these studies, the quenching is thus related to the temperature or the density of the gas, i.e., to the gas pressure around the Cosmic Web.

In \cite{tanimura_fil}, we have studied the hot gas via the SZ emission around cosmic filaments detected with DisPerSE on SDSS galaxies in the range $0.2<z<0.6$. The temperature of the gas around the filaments has been estimated to be $\mathrm{T}\sim10^6$ K (larger than the temperature of quenching of the halos found in \cite{gabor2015}: $\mathrm{T} \geq 10^{5.4}$ K). 

The quiescent fraction profile measured in the previous section and shown in Fig.~\ref{chapt:filaments;red_over_blue} might depend mostly on the processes of quenching, that drives the population changeover. We have modelled in a simple way the $f_\mathrm{Q}$ profile with a model of gas density distribution around the cosmic filaments. We have used a $\beta$-model (historically used to model the gas density profiles in galaxy clusters \citep{cavaliere1978}). It writes as:

\begin{equation}\label{chapt:filaments;eq:beta}
f_\mathrm{Q}(r) = \frac{{f_\mathrm{Q}}_0}{{\left(1+{\left(\frac{r}{r_\mathrm{s}}\right)}^2\right)}^{\frac{3}{2}\beta}} + c,
\end{equation}

where ${f_\mathrm{Q}}_0$ is the mean ratio of excess of passive galaxies over the sum of the excess of active and passive galaxies in the center of the filaments, $r_\mathrm{s}$ is the core radius, $\beta$ the slope of the profile, and $c$ is the background value.

\begin{figure}[!ht]
\centering
\includegraphics[width=0.5\textwidth]{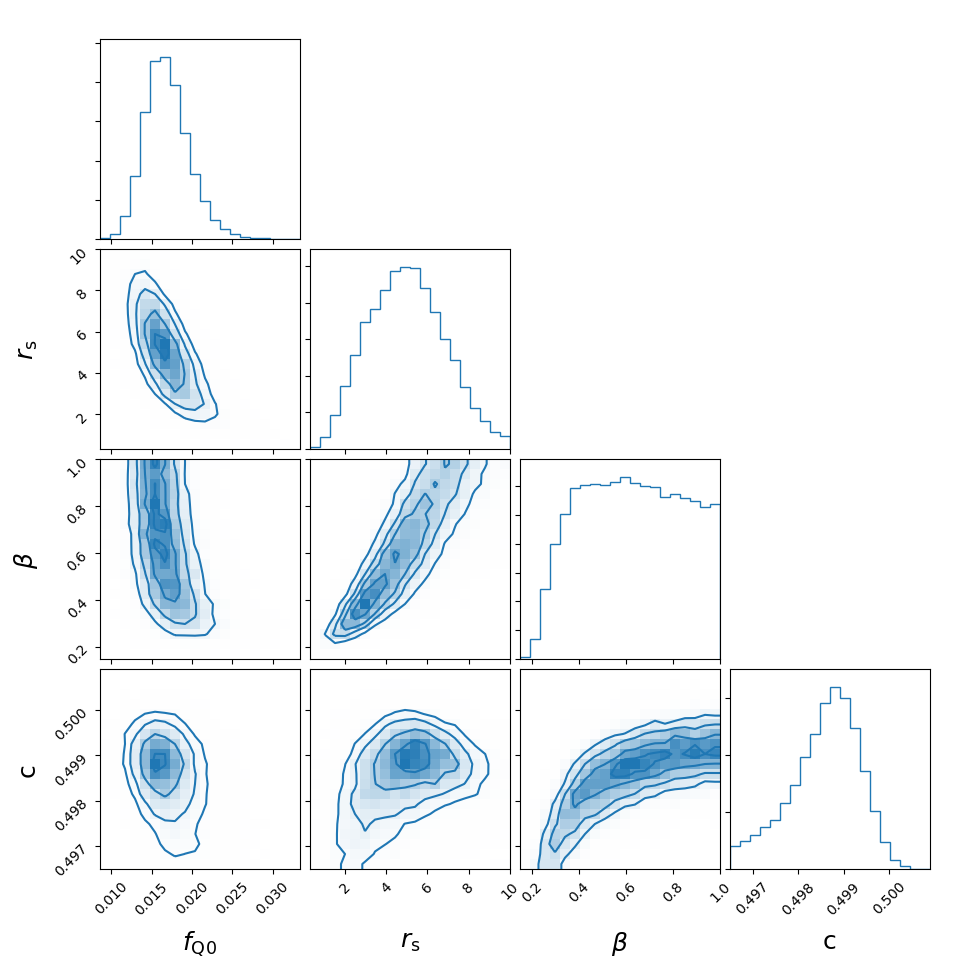}
\caption{\label{chapt:filaments;mcmc}Posterior distributions of the four parameters of the $\beta$-model (Eq.~\ref{chapt:filaments;eq:beta}), fitted to the quiescent fraction profile shown in Fig.~\ref{chapt:filaments;red_over_blue}.}
\end{figure}

We have performed an MCMC analysis and obtained the posterior distributions of the four parameters of Eq.~\ref{chapt:filaments;eq:beta}. The distributions and the correlations are shown in Fig.~\ref{chapt:filaments;mcmc}. In Fig.~\ref{chapt:filaments;red_over_blue} is also displayed in orange 1000 models randomly picked from the MCMC distribution. The median parameters are: ${f_\mathrm{Q}}_0 = 0.017\pm0.003$, $r_\mathrm{s} = 4.4\pm1.7$ Mpc, $\beta = 0.54\pm0.18$, and the background value $c=0.498\pm0.001$. Assuming the quiescent fraction of galaxies traces the pressure of the gas responsible of the quenching, the values of the slope $\beta$ of the gas profile and of the quiescent fraction profile should be the same. The slope $\beta$ is not well constrained as seen in the distribution in Fig.~\ref{chapt:filaments;mcmc}: it is fully degenerate with the parameter $r_\mathrm{s}$. However, $\beta=2/3$ which is the case for projected iso-thermal gas in hydrostatic equilibrium (scenario supported by numerical simulation, e.g., in \cite{gheller2019}), is encompassed in the values allowed by the MCMC. Moreover, the value $\beta=2/3$ that fits the gas profile in SZ in \cite{tanimura_fil} is also encompassed in the allowed values by the MCMC.

\section{Discussion and Summary}\label{chapt:filaments;sect:conclusion}

We have studied in detail the statistical properties of the galaxies from the WISExSCOS value-added catalogue around cosmic filaments detected with DisPerSE in the SDSS. We have measured with a high significance ($\gtrsim 5 \sigma$) galaxy over-density radial profiles on the full sample of galaxies, and also on the three populations: active, transitioning, and passive galaxies.

Despite some biases on the measurement due to the methodology or to the data themselves, we have fitted an average profile of galaxy over-density around cosmic filaments with an exponential law. We have obtained a typical radius of $r_\mathrm{m} = 7.5\pm0.2$ Mpc. We have also pointed out the evidence of a higher excess of passive galaxies than transitioning galaxies, and a higher excess of transitioning galaxies than active galaxies near the filament's spines. This excess of passive galaxies induces an SFR and an $\mathrm{M}_*$ gradient pointing towards the filament's spines, that we have also detected. This indicates that there are more passive galaxies near the filament's spines, in agreement with the previous studies \citep[e.g.,][]{malavasi2017, kraljic2018, laigle2018, sarron2019}.

We have also studied the excess of $\mathrm{M}_*$ and of SFR for the three galaxy populations, and have pointed out the evidence of a positive $\mathrm{M}_*$ gradient for the active, the transitioning, and the passive populations of galaxies. This means that galaxies are more massive in the filaments, whatever their types, in agreement with \cite{kraljic2018} and \cite{malavasi2017} who observed a stellar mass gradient for star forming galaxies. We have also detected a positive SFR gradient for passive galaxies, showing that passive galaxies are more star-forming in the filaments. These gradients could be related to galaxy merges inside filaments, that increase the $\mathrm{M}_*$ of the galaxies and may rejuvenate the SFR in passive galaxies.

We have investigated how the quiescent fraction of galaxies, $f_\mathrm{Q}$, behaves around cosmic filaments. Following previous studies showing the role of the Cosmic Web and of the gas in the quenching of star formation, and following the recent detection of the diffuse hot gas around cosmic filaments in SZ with $\mathrm{T}\sim10^6$ K, we have modelled in an exploratory way the $f_\mathrm{Q}$ profile with a model of gas density distribution. We have measured parameters with an MCMC analysis. The obtained slope $\beta = 0.54\pm0.18$ is not well constrained. However, the MCMC distribution encompasses the value $\beta=2/3$ that is the case for projected iso-thermal gas in hydrostatic equilibirum, that is suggested for example by \cite{gheller2019}. The MCMC distribution of $\beta$ also encompasses the values of $\beta$ fitted on gas profile with SZ. The slope of the quiescent fraction profile is not inconsistent with the slope of a gas profile around cosmic filaments. Therefore, in this exploratory work, it is not excluded that the gas around cosmic filaments might have a non-negligible role in the quenching of star formation in galaxies.

To capture and understand the connections between the Cosmic Web environment and the galaxy properties, the profiles derived and shown in this study will be compared with numerical simulations in Illustris-TNG (Gal\'arraga et al., in prep.) and with observations of the gas around cosmic filaments (with the SZ effect in \textit{Planck}, or in the X-rays with the future e-Rosita\footnote{\url{https://www.mpe.mpg.de/eROSITA}} mission). In the context of future very large galaxy surveys like Euclid\footnote{\url{https://www.euclid-ec.org}}, LSST\footnote{\url{https://www.lsst.org}}, or WFIRST\footnote{\url{https://www.nasa.gov/wfirst}}, such studies on galaxy properties will be possible for a broader range of redshift, for a wider field of view, and thus with a higher significance.

\appendix

\section{Masking the galaxy cluster's members}\label{appendix:mask}

\begin{figure*}[!ht]
\centering
\includegraphics[width=\textwidth]{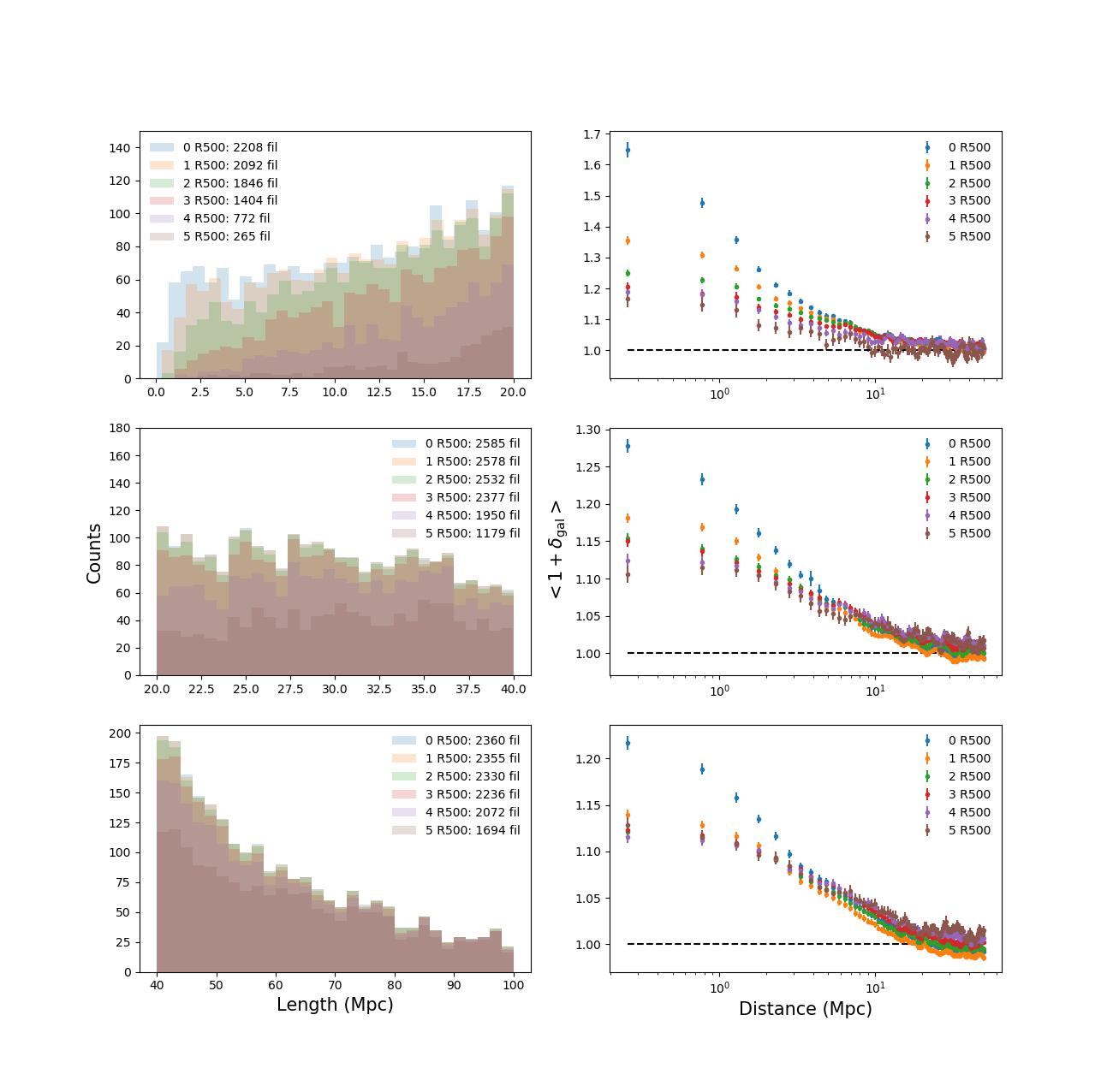}
\caption{\label{chapt:filaments;compare_r500}Comparison of the effect of different masks. Galaxy clusters with $z<0.4$ from the PSZ2, MCXC, RedMaPPer, AMF9, WHL12 and WHL15 catalogues are masked with areas of radii from 0 to $5\times\mathrm{R}_{500}$. The left and the right columns show the histogram of filament's lengths and the over-density profiles $<1+\delta_\mathrm{gal}>$ respectively, for the short, regular, and long filaments (from top to bottom). We have chosen the mask at $3\times\mathrm{R}_{500}$.}
\end{figure*}

\begin{figure*}[!ht]
\centering
\includegraphics[width=\textwidth]{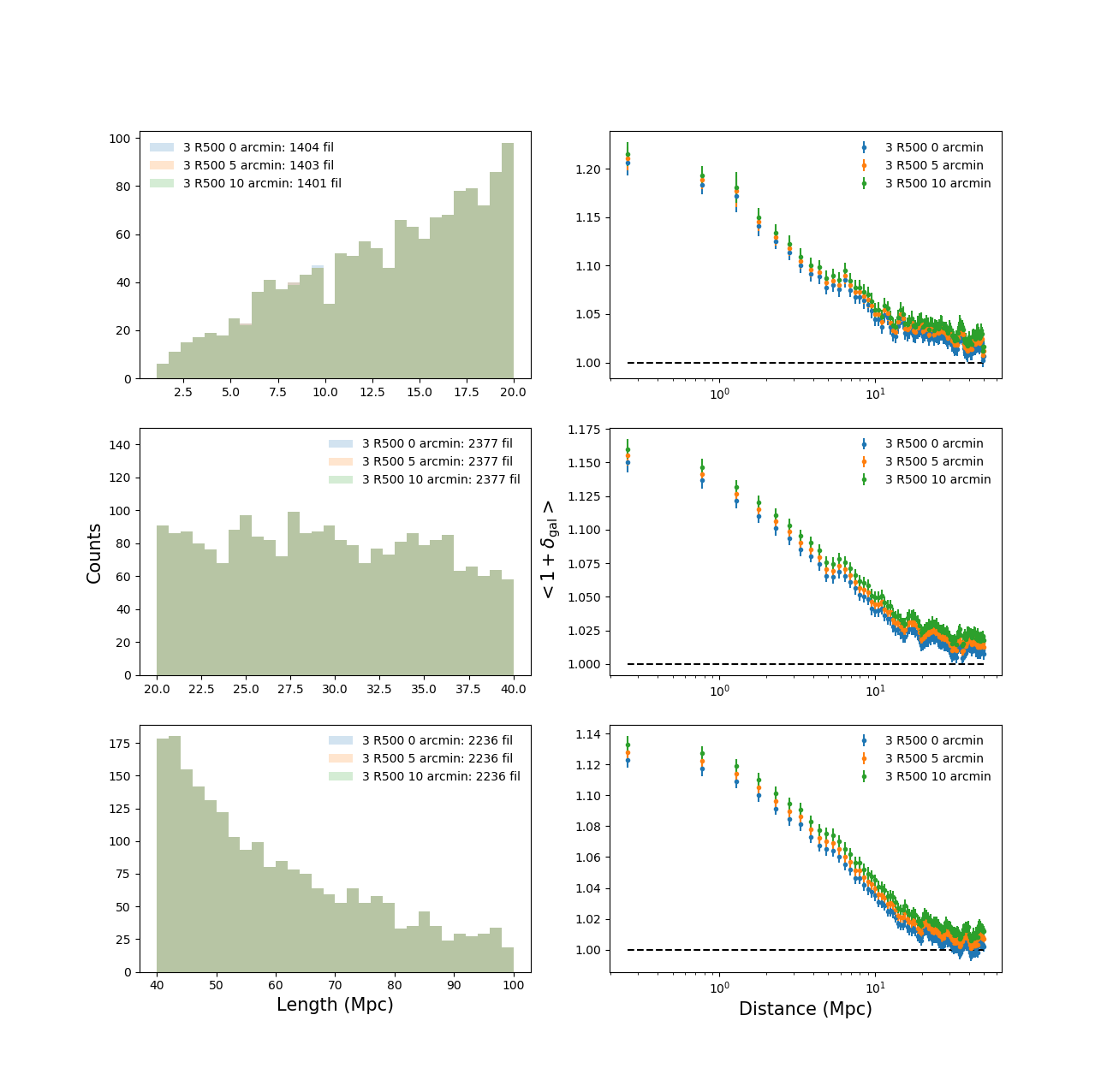}
\caption{\label{chapt:filaments;compare_arcmin}Comparison of the effect of different masks. For the galaxy clusters without estimated radius in the \textit{Planck} PSZ2 catalogue with $z<0.4$, regions defined by areas of 0, 5, and 10 arcmin radii are masked. No difference is noticed and all plots are overlapping. To clearly see the different profiles, we have added +0.1 and +0.2 to the y axis for the orange line and the green line, respectively. We have therefore chosen to mask at 5 arcmin.}
\end{figure*}

\begin{figure*}[!ht]
\centering
\includegraphics[width=\textwidth]{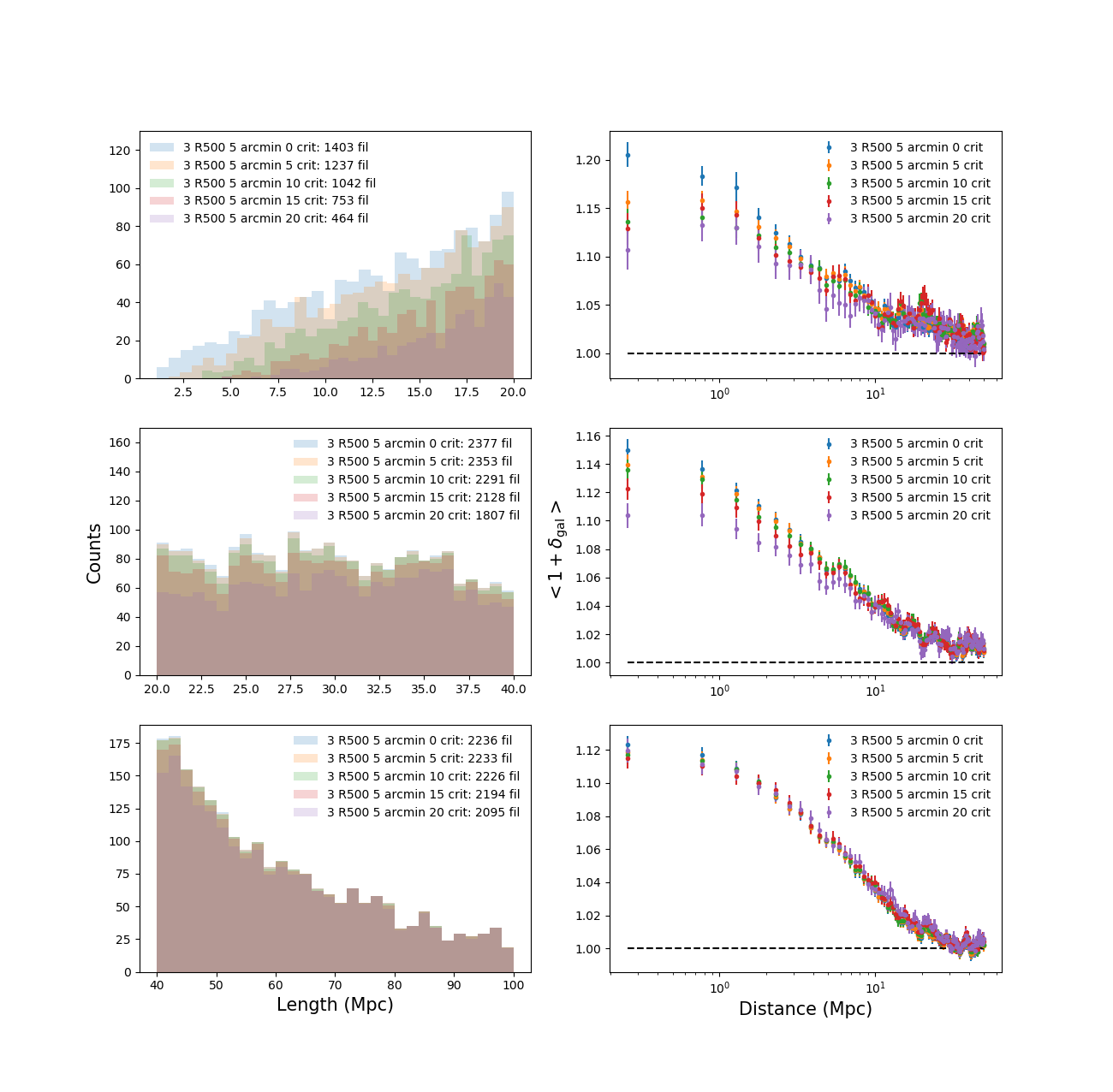}
\caption{\label{chapt:filaments;compare_crit}Comparison of the effect of different masks. Two type of critical points given by DisPerSE are masked: the maxima density points and the bifurcation points. The regions around the critical points with $z<0.4$ are masked from 0 to 20 arcmin. Right panel, the signal is decreasing with the size of the masks for short and regular filaments. Left panel, the masks remove the short filaments.}
\end{figure*}

\begin{acknowledgements}The authors thank useful discussions with Guillaume Hurier, and with the members of the ByoPiC project (\url{https://byopic.eu/team}. This research has been supported by the funding for the ByoPiC project from the European Research Council (ERC) under the European Union's Horizon 2020 research and innovation programme grant agreement ERC-2015-AdG 695561. This publication made use of the SZ-Cluster Database (\url{http://szcluster-db.ias.u-psud.fr}) operated by the Integrated Data and Operation Centre (IDOC) at the Institut d'Astrophysique Spatiale (IAS) under contract with CNES and CNRS. This research has made use of data obtained from the SuperCOSMOS Science Archive, prepared and hosted by the Wide Field Astronomy Unit, Institute for Astronomy, University of Edinburgh, which is funded by the UK Science and Technology Facilities Council.
\end{acknowledgements}

\bibliographystyle{aa}
\bibliography{biblio}

\end{document}